%% file: 0sigconf.tex
\documentclass[sigconf]{acmart}
\AtBeginDocument{%
  }

\copyrightyear{2026}
\acmYear{2026}
\setcopyright{cc}
\setcctype{by}
\acmConference[SIGIR '26]{Proceedings of the 49th International ACM SIGIR Conference on Research and Development in Information Retrieval}{July 20--24, 2026}{Melbourne, VIC, Australia}
\acmBooktitle{Proceedings of the 49th International ACM SIGIR Conference on Research and Development in Information Retrieval (SIGIR '26), July 20--24, 2026, Melbourne, VIC, Australia}
\acmDOI{10.1145/3805712.3808378}
\acmISBN{979-8-4007-2599-9/2026/07}

\usepackage{subcaption}
\usepackage{graphicx}
\usepackage{tikz}
\usetikzlibrary{shapes, arrows, positioning, fit, backgrounds}
\usepackage{enumitem}
\usetikzlibrary{calc}

\begin{document}

\title{H-MAPS: Hierarchical Memory-Augmented Proactive Search Assistant for Scientific Literature}

\author{Koji Nishikawa}
\email{s2111118@klis.tsukuba.ac.jp}
\affiliation{%
  \institution{University of Tsukuba}
  \city{Tsukuba}
  \country{Japan}
}

\author{Makoto P. Kato}
\email{mpkato@acm.org}
\affiliation{%
  \institution{University of Tsukuba}
  \city{Tsukuba}
  \country{Japan}
  }
\affiliation{  
  \institution{National Institute of Informatics}
  \city{Tokyo}
  \country{Japan}  
  }


\begin{abstract}
    Scientific reading is an active process that frequently requires consulting external resources, but manual keyword searching interrupts the reading flow and imposes a high cognitive load. Existing proactive information retrieval systems often suffer from context ambiguity, as they rely solely on on-screen text and ignore the reader's specific background and intent. 
    In this demonstration, we present H-MAPS ({\bf H}ierarchical {\bf M}emory-{\bf A}ugmented {\bf P}roactive {\bf S}earch Assistant), a proactive literature exploration assistant that resolves this ambiguity by leveraging a three-layered hierarchical memory. Triggered by implicit reading behaviors, H-MAPS articulates the user's latent information needs into explicit natural language questions and performs neural retrieval entirely on the local device to ensure privacy.
    We demonstrate H-MAPS using a scenario where two researchers, specializing in NLP and HCI, read the same paper. In response, the system generates profile-specific questions and retrieves distinct literature tailored to each user.
\end{abstract}





\maketitle

\input{1introduction}

\input{2RelatedWork}
\input{3H-MAPS_Architecture}
\input{4Demo}
\input{5Conclusion}
\input{6Acknowledgments}

\bibliographystyle{ACM-Reference-Format}
\bibliography{sample-base}

\appendix

\end{document}

%% file: 1introduction.tex
\section{Introduction}
\label{sec:introduction}

Scientific reading is an active process of integrating new information with existing knowledge. During this process, researchers frequently need to consult external resources to verify definitions, explore related work, or contextualize citations~\cite{Koskela2019capture}. However, this workflow suffers from two major challenges.

First, the cognitive load of manual search is high. Upon encountering an information need, the reader must interrupt their deep thought process to formulate a query, execute a search, and filter through results. This explicit task switching can disrupt the reading flow and hinder deep engagement~\cite{Koskela2019capture}.
Second, existing automated solutions suffer from context ambiguity. While proactive information retrieval systems attempt to present information automatically, most rely solely on on-screen text as search keywords~\cite{vuong2017proactive}. This approach overlooks the critical context of \emph{who} is reading and \emph{why}, failing to account for the significant variation in what different individuals consider relevant to the same query~\cite{teevan2010potential}. For instance, when reading about ``Retrieval-Augmented Generation (RAG),'' an NLP researcher might seek details on computational latency, whereas an HCI researcher might focus on user trust. Ignoring the reader's background, specifically the interaction between their long-term interests and short-term session behavior~\cite{bennett2012modeling}, inevitably leads to irrelevant, noisy retrieval results.

To address these challenges, we present \textbf{H-MAPS} ({\bf H}ierarchical {\bf M}emory-{\bf A}ugmented {\bf P}roactive {\bf S}earch Assistant), a proactive literature search assistant\footnote{A video demonstration of the system is available at \url{https://qr1.jp/5A3icZ}}.
To address the disruption of manual search, H-MAPS enables a {\it query-free} experience by automatically detecting information needs without requiring explicit user input. To resolve context ambiguity, the system utilizes a Hierarchical Memory composed of short- and long-term reading contexts and an inferred user profile, allowing it to deeply understand \emph{who} is reading. Based on this memory, H-MAPS formulates explicit questions to guide the search and presents results paired with these questions, thereby clarifying \emph{why} specific information is relevant to the user's intent.

This demonstration paper makes the following contributions:
\begin{itemize}[leftmargin=*]
\item {\bf Proactive Support for Scientific Reading:} We propose a proactive search assistant that minimizes cognitive load by automating retrieval, enabling access to external resources without interrupting the reading flow or formulating queries.

\item {\bf Context-Aware Retrieval via Hierarchical Memory:} We introduce a Hierarchical Memory  synthesizing short-term context, long-term interests, and inferred profiles to resolve ambiguity and retrieve personalized literature tailored to \emph{who} is reading.

\item {\bf Explainable Presentation via Question Articulation:} We demonstrate an explainable interface that verbalizes latent needs into explicit natural language questions, providing a clear rationale for \emph{why} results are recommended to facilitate verification.

\end{itemize}

%% file: 2RelatedWork.tex
\section{Related Work}
\label{sec:related}

\textbf{Proactive Search and Screen Surveillance.} 
Proactive search systems aim to automatically retrieve relevant information by anticipating user needs based on their primary task context. 
Prior works have explored utilizing pre-search contexts, such as recently browsed web pages~\cite{kong2015predicting}, or capturing text from the primary task window (e.g., writing an essay) to formulate background queries~\cite{liebling2012anticipatory, Koskela2019capture}. 
To generalize this approach across diverse desktop applications, screen surveillance techniques have been proposed. For example, Vuong et al. continuously monitored screen content via optical character recognition (OCR) to model users' topical activities and proactively retrieve relevant documents~\cite{vuong2017proactive, vuong2017watching}. 
While these systems successfully alleviate the burden of manual query formulation, they predominantly rely on extracting keywords directly from the on-screen text. This often leads to context ambiguity, as isolated keywords fail to capture the deeper semantics of the task or the user's specific cognitive intent.

\textbf{Personalized Search and Context Modeling.} 
Extensive research has demonstrated that search intent varies significantly among individuals; different users often find completely different results relevant to the identical query~\cite{teevan2010potential}. 
Traditional personalization strategies often rely on server-side query logs and session histories to infer user interests~\cite{bennett2012modeling}. However, such centralized logging approaches raise significant user privacy concerns~\cite{reimer2023archive}.
Recently, there is a growing trend of leveraging Large Language Models (LLMs) to automatically generate search queries based on user interaction history~\cite{ouyang2025token}. Furthermore, for proactive search scenarios, translating noisy contexts into concise queries has proven effective in bridging the input gap for off-the-shelf retrievers~\cite{meng2025bridging}.
Building upon these foundations, H-MAPS integrates screen surveillance with a hierarchical memory structure in a fully privacy-preserving, on-device manner.
Unlike existing proactive reading assistants, H-MAPS does not merely use raw on-screen text as search queries. Instead, it dynamically synthesizes local context, session context, and long-term user profiles using local LLMs to articulate latent needs into explicit natural language questions. This approach not only provides highly personalized retrieval but also offers an explainable rationale for the recommendations, effectively resolving context ambiguity without compromising privacy.

%% file: 3H-MAPS_Architecture.tex
\section{H-MAPS Architecture}
\label{sec:system}

\setcounter{figure}{1}
\begin{figure*}[t]
  \centering
  \includegraphics[width=0.92\textwidth]{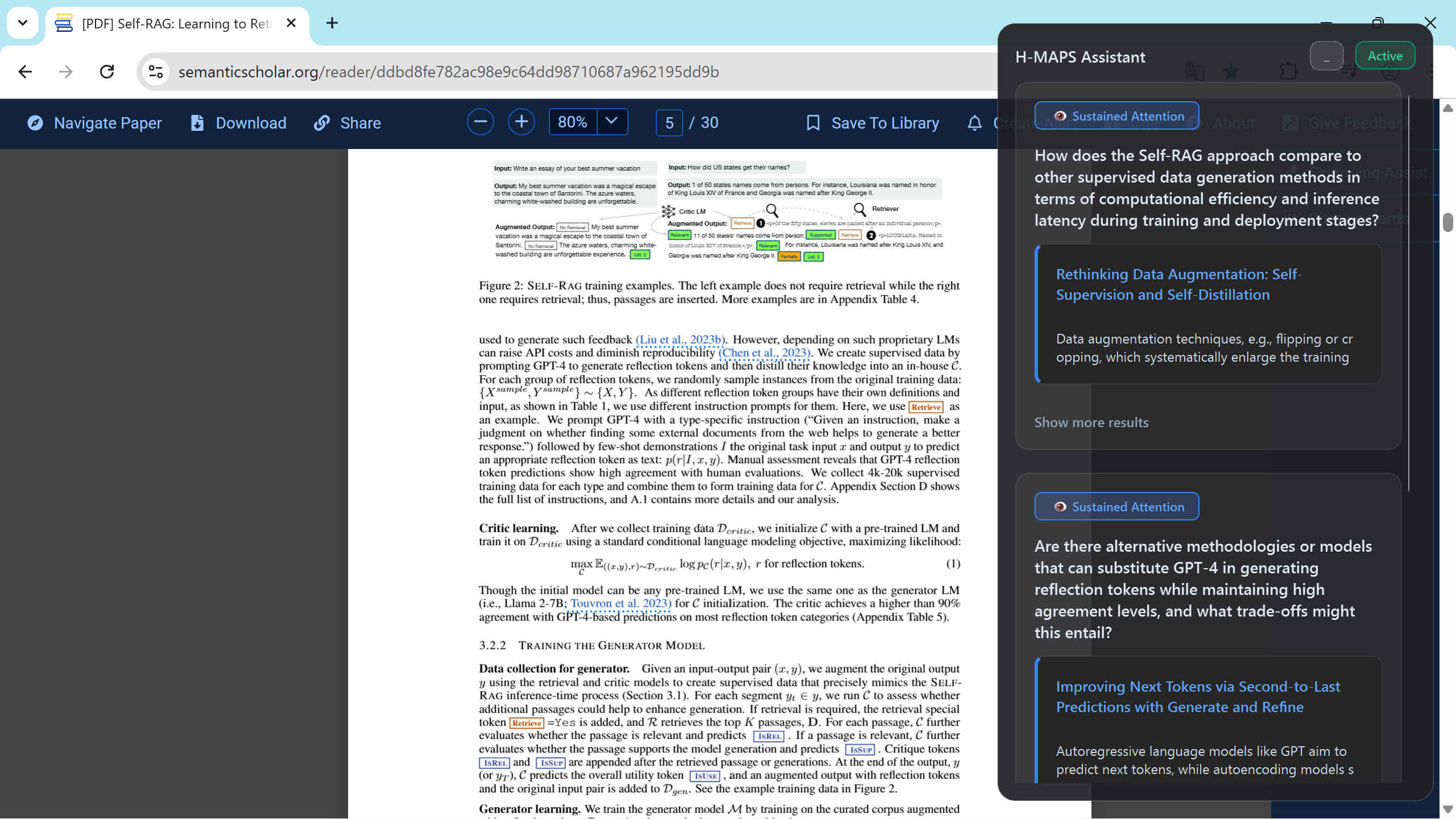}
  \caption{H-MAPS overlay UI. The assistant operates as a peripheral overlay on the desktop, generating multiple literature search questions in response to the behavior-driven trigger and displaying the top papers found for each question. The search panel on the right is enlarged for clarity.}
  \label{fig:ui_overview}
\end{figure*}

\setcounter{figure}{0}
\begin{figure}[t]
  \centering
  \includegraphics[width=\linewidth]{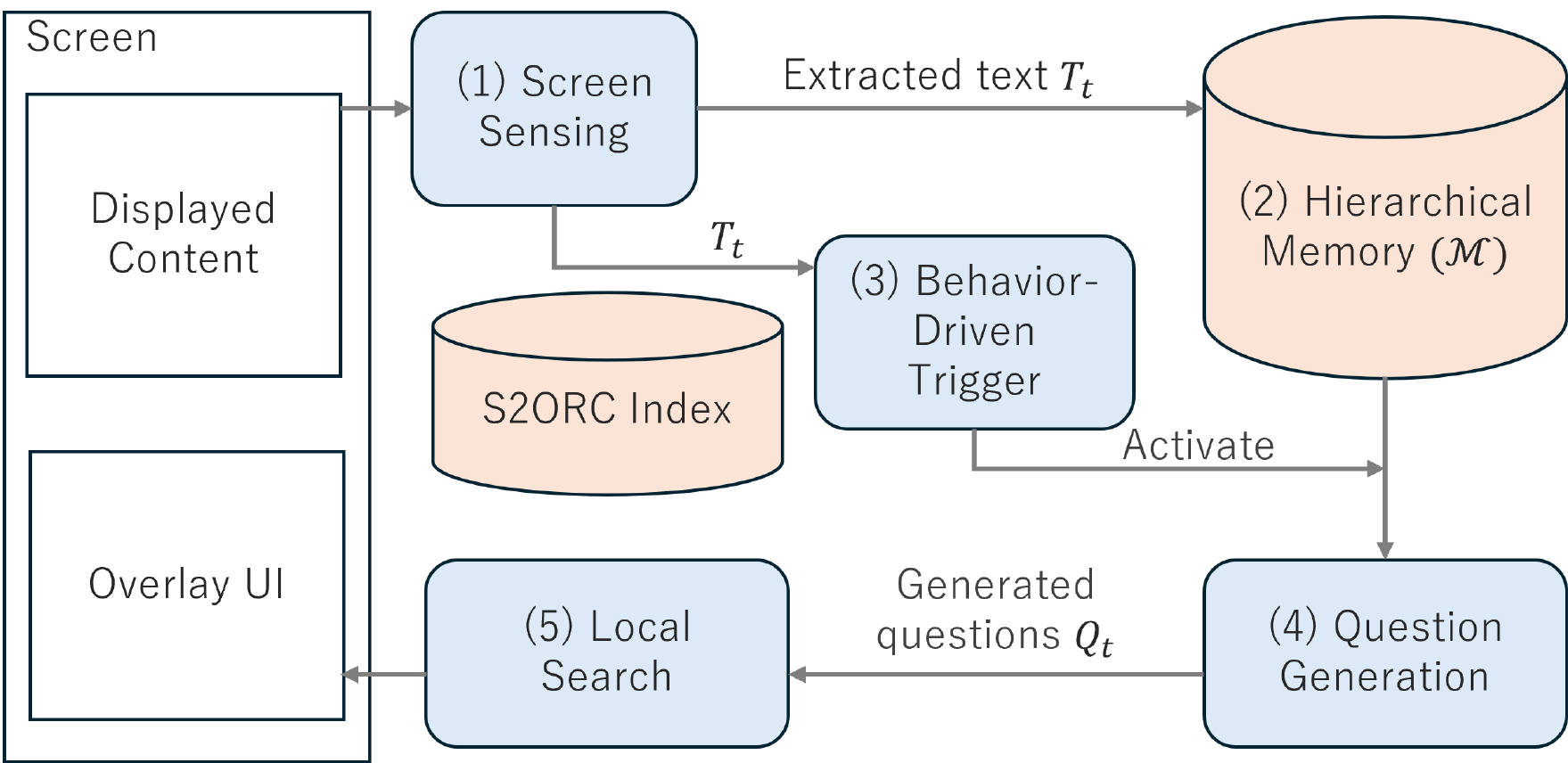}
  \caption{System architecture of H-MAPS, which comprises five key components: (1) Screen Sensing captures desktop content; (2) Hierarchical Memory maintains context; (3) Behavior-Driven Trigger detects intervention moments; (4) Question Generation articulates information needs; and (5) Local Search performs neural retrieval and presents search results in the Overlay UI.}
  \label{fig:architecture}
\end{figure}

\setcounter{figure}{2}

As shown in Figure~\ref{fig:architecture},
H-MAPS continuously captures screen content and applies OCR to populate a hierarchical memory with short- and long-term contexts and the user profile. Triggered by specific user behaviors, H-MAPS articulates the user's latent information needs, conducts a literature search, and presents the results via a screen overlay.
The architecture consists of five components: (1) Screen Sensing, (2) Hierarchical Memory, (3) Behavior-Driven Triggers, (4) Question Generation, and (5) Local Search. 
Figure~\ref{fig:ui_overview} illustrates the H-MAPS interface as a user reads a RAG paper, with retrieved literature displayed on the right side of the screen.
The implementation details, including prompts used for LLMs, can be found at our code repository\footnote{\url{https://github.com/kasys-lab/H-MAPS} \label{fn:code}}.

\subsection{Screen Sensing}
To support diverse reading environments without relying on specific software integrations, H-MAPS visually monitors the desktop.
We employ the \texttt{mss} library for high-speed, cross-platform screen capture.
Text extraction is performed entirely on-device using the Tesseract OCR engine via the \texttt{pytesseract} wrapper, running at approximately 1\,Hz to capture reading dynamics.
The extracted text at time $t$ is denoted by $T_t$ in Figure~\ref{fig:architecture}.

Since scientific reading may involve unpublished or sensitive content, raw screen captures and OCR text are processed and stored only on-device.
Furthermore, we keep capture, OCR, memory, and retrieval fully local. 
As an additional safeguard against unexpected data leak, we summarize OCR locally with a lightweight model (\texttt{Phi-3.5-mini}) and remove identifiers (names, URLs, email addresses, and numbers). 

\subsection{Hierarchical Memory}
Scientific reading involves aligning immediate focus with long-term research goals.
To support this cognitive process, H-MAPS constructs a three-layered hierarchical memory $\mathcal{M} = \langle M^{\text{loc}}, M^{\text{ses}}, m^{\text{prof}} \rangle$. 
To balance personalization with privacy, all layers are synthesized entirely on-device using a lightweight local LLM (denoted as $f_{\text{LLM}}$). 
Note that the specific numerical values in parentheses throughout this section indicate the parameters used in our actual implementation.

\textbf{Local Context ($M^{\text{loc}}$):} 
Represents the specific reading context of the last few minutes. Given the extracted text $T_t$, the system extracts the newly appeared text lines $\Delta T_t$ from the previous frame and appends them to an internal buffer $B$. When the buffer size $|B|$ exceeds a threshold $\theta_{\text{loc}}$ (2,000 characters), the local LLM generates a sanitized micro-summary:
\begin{equation}
m^{\text{loc}} = f_{\text{LLM}}(p_s, B)
\end{equation}
where $p_s$ is a prompt instructing the LLM to summarize the text. This summary $m^{\text{loc}}$ is then pushed to the history stack $M^{\text{loc}}$, and the buffer $B$ is cleared.

\textbf{Session Context ($M^{\text{ses}}$):} 
Maintains a running summary of the document read during the current session. A background thread periodically (every $\theta_{\text{ses}} = 300$ seconds) retrieves the $k_{\text{loc}}$ most recent entries from the stack $M^{\text{loc}}$ ($k_{\text{loc}} = 10$) and synthesizes them to capture the paper's logical flow and the knowledge accumulated so far:
\begin{equation}
m^{\text{ses}} = f_{\text{LLM}}(p_i, M^{\text{loc}}_{1 \ldots k_{\text{loc}}})
\end{equation}
where $p_i$ is a prompt instructing the LLM to integrate the multiple context items. This summary $m^{\text{ses}}$ is then pushed to the session history stack $M^{\text{ses}}$. This process creates a coherent understanding of the current task, rather than a mere concatenation of fragmented texts.

\textbf{Inferred Profile ($m^{\text{prof}}$):} 
Describes the user's long-term research interests as a single natural language text. To reflect gradual topic shifts and concept drift without losing past context, $m^{\text{prof}}$ is asynchronously updated whenever a new $m^{\text{ses}}$ is generated:
\begin{equation}
m^{\text{prof}} \leftarrow f_{\text{LLM}}(p_u, \{m^{\text{prof}}\} \cup M^{\text{ses}}_{1 \ldots k_{\text{ses}}})
\end{equation}
where $M^{\text{ses}}_{1 \ldots k_{\text{ses}}}$ denotes the $k_{\text{ses}}$ most recent session summaries ($k_{\text{ses}} = 5$ in our setup), and $p_u$ is a prompt instructing to update the profile based on the new session. By injecting this hierarchical memory $\mathcal{M}$ into the prompt, H-MAPS achieves highly personalized retrieval support based on \emph{what is currently being viewed} ($M^{\text{loc}}$), \emph{in what context} ($M^{\text{ses}}$), and \emph{who} is reading ($m^{\text{prof}}$).

\subsection{Behavior-Driven Trigger}
H-MAPS uses two simple heuristics derived from OCR-text dynamics to decide when to present proactive suggestions.
To quantify screen similarity, we compute the Jaccard similarity $J(T_t, T_{t-\Delta})$ based on the set of character bi-grams extracted from the text $T_t$, which is robust to minor OCR or layout noise.
Importantly, these triggers are designed to capture distinct cognitive intents rather than merely measuring scroll distances.

\textbf{Sustained Attention.}
This trigger identifies deep engagement when a user pauses scrolling, indicating an intent to \emph{explore and deeply understand} new concepts. To distinguish reflection from mere reading of dense passages, we employ an adaptive threshold:
$\theta = \max(10, \min (60, |T_t| / v ))$
where $v$ is the reading speed (char/sec) estimated via regression of scroll intervals.
The trigger fires when $J(T_t, T_{t-\Delta}) > 0.9$ persists for $\Delta > \theta$.

\textbf{Content Revisit.}
This trigger detects when a user scrolls back, indicating an intent to \emph{clarify forgotten context or verify past definitions}. We maintain a rolling history buffer $\mathcal{H}_t$ of screen states from the last 180 seconds. 
A revisit event is detected when the current screen text $T_t$ exhibits high similarity to a past state within $\mathcal{H}_t$:
$\exists T \in \mathcal{H}_t \text{ s.t. } J(T_t, T) > 0.8$.

\subsection{Question Generation}
A distinctive feature of H-MAPS is that instead of using on-screen text directly as search queries, it translates the user's latent information needs into explicit information needs formulated as natural language questions, denoted by $Q_t$ in Figure~\ref{fig:architecture}. This step shifts the user's cognitive task from \emph{inferring} why a specific search result is useful to simply \emph{verifying} whether the system-presented question matches their intent. 
This makes the system's intent inspectable: users can quickly accept or dismiss suggestions by checking whether the search question matches what they want, rather than inferring intent only from a ranked list of paper titles. This transition from inference to verification can reduce the cognitive load required to assess the relevance of retrieved documents.

To accurately articulate these needs, the system dynamically retrieves relevant summaries from the hierarchical memory.
Specifically, each summary in the Local and Session Contexts is pre-encoded into a vector representation using a dense retrieval model (\texttt{intfloat/e5-small-v2}~\cite{wang2022text} in our implementation) upon creation and stored in an on-device cache.
At query time, the system encodes the current on-screen text $T_t$ and performs a nearest neighbor search against these cached memory embeddings.
We retrieve the top-2 most relevant summaries from each layer and concatenate them with the user's Inferred Profile ($m^{\text{prof}}$) to construct the prompt for the LLM. 
The system feeds these memory contexts as input to generate explicit information needs that align with the user's momentary cognitive focus. Critically, the prompt of question generation is dynamically conditioned by the detected behavioral trigger. When {\bf Sustained Attention} is detected, the system interprets this behavior as a sign of deep engagement. Consequently, it instructs the model to formulate questions that facilitate exploration: broadening the scope by inquiring about related work, alternative methodologies, or critical limitations of the concept being read. Conversely, when {\bf Content Revisit} is detected, the system interprets this scroll-back behavior as a sign that the user is attempting to recall a forgotten concept. In this case, it instructs the model to formulate questions that facilitate clarification: focusing on factual verification, such as definitions of specific terms or summaries of previously introduced concepts. 
The specific prompts used for question generation are available in our code repository$^{\text{\ref{fn:code}}}$.

\subsection{Local Search}
To identify relevant literature for the natural language questions generated in the previous section, H-MAPS executes a fully local retrieval pipeline.
We adopt a dense retrieval approach to capture contextual semantic similarity.
General-purpose retrieval models without domain-specific fine-tuning often face a semantic gap between long, highly specific natural language questions and brief paper abstracts.
To bridge this gap and capture contextual semantic similarity, we fine-tuned \texttt{intfloat/e5-small-v2} on the LitSearch dataset~\cite{ajith2024litsearch}, which provides domain-specific pairs of 597 scientific information-seeking queries and relevant papers.
Specifically, we trained the bi-encoder model using the Tevatron framework with InfoNCE loss, utilizing in-batch negatives for contrastive learning. 
Following the e5 model specifications, we prepended ``query: '' and ``passage: '' to the inputs and applied mean pooling with L2 normalization. 
The model was fine-tuned for 1 epoch with a learning rate of $2 \times 10^{-5}$, a batch size of 16, a temperature of 0.05, and a maximum sequence length of 512 tokens.
In our preliminary evaluation on the held-out LitSearch test queries, this domain-specific fine-tuning substantially improved retrieval accuracy, increasing the MRR@10 from 0.273 to 0.358 compared to the baseline.

For the search corpus, we utilize the S2ORC dataset~\cite{Lo2020s2orc} covering a wide range of scientific fields.
To ensure retrieval quality, we filter out papers without abstracts, resulting in an index of approximately 14 million papers.
Each paper is pre-encoded by concatenating its title and abstract, and stored in a local FAISS index~\cite{johnson2021billion}, an efficient library for dense similarity search.

The system performs nearest neighbor search for each question against the FAISS index.
Based on the retrieved document IDs, metadata (titles, authors, and URLs) is immediately resolved from a co-located SQLite database.
This on-device configuration avoids transmitting user queries and retrieved results to external servers and reduces network latency.
In our implementation, the index and metadata database require approximately 20.5GB and 50.1GB, respectively, to cover 14 million papers. 
This compact footprint demonstrates the feasibility of hosting the dense retrieval index entirely on a local device, allowing users to perform offline retrieval and periodically update the corpus.

%% file: 4Demo.tex
\section{Demonstration Scenario}
\label{sec:demo}

\begin{table*}[t]
  \caption{Representative output logs from the demo scenario. Despite viewing identical content (Page 5 of the Self-RAG paper), the system generated profile-specific questions and retrieved distinct literature from the S2ORC corpus.}
  \label{tab:demo_results}
  \small
  \begin{tabular}{p{0.08\linewidth} p{0.38\linewidth} p{0.48\linewidth}}
    \toprule
    \textbf{Profile} & \textbf{Generated Question} & \textbf{Top-Retrieved Papers} \\
    \midrule
    \textbf{NLP} & What are recent studies or papers that explore \textbf{inference latency optimization} in Retrieval-Augmented Generation (RAG) models enhanced with reflection tokens, particularly focusing on Conditional Language Models trained with supervision or GPT-4 distilled prompts? & 
\begin{tabular}[t]{lp{0.85\linewidth}}
    1. &\textbf{FedRAG}: A Framework for Fine-Tuning RAG Systems \\
    2. & \textbf{RAG Foundry}: A Framework for Enhancing LLMs for Retrieval Augmented Generation
\end{tabular}\\    
    \midrule
    \textbf{HCI} & What studies explore methods for incorporating reflection tokens in Human-AI collaboration systems to enhance \textbf{trust and transparency} in model outputs? & 
\begin{tabular}[t]{lp{0.85\linewidth}}
    1. & \textbf{Designing Transparency} for Effective Human-AI Collaboration \\
    2. & Investigating the Relationship between AI and Trust in Human-AI Collaboration
\end{tabular}\\
    \bottomrule
  \end{tabular}
\end{table*}
To demonstrate the effectiveness of H-MAPS in resolving the ambiguity of on-screen content, we present a scenario involving two researchers with contrasting expertise reading the same scientific paper: \textit{Self-RAG}~\cite{asai2024selfrag}. This demonstration highlights how the system leverages hierarchical memory to transform identical visual stimuli into personalized retrieval support.

\subsection{User Profiles and Reading Context} 
We initialized the system with two distinct Inferred Profiles.
\textbf{NLP profile:} a user is an NLP researcher specializing in RAG algorithms.
Primary interests include inference latency optimization and the computational complexity of reflection token generation.
The user prioritizes backend performance and system-level efficiency over user interface design.
\textbf{HCI profile:} a user is an HCI researcher focusing on Human-AI collaboration and trust calibration.
Primary interests include the impact of reflection tokens on user cognitive load and interface transparency.
The user prioritizes explaining AI outputs over backend algorithmic details.

The scenario begins with both researchers reading a paper at a steady pace of approximately 100 characters per second. During this process, H-MAPS continuously monitors the screen and accumulates a Local Context in its hierarchical memory, capturing key concepts described in the paper, such as the definition of ``reflection tokens'' and the overall Self-RAG framework.

\subsection{Triggering and Question Generation} 
A critical intersection occurs at Page 5 of the Self-RAG paper, where ``Training the Critic Model'' is explained in the paper (see Figure~\ref{fig:ui_overview}). This section describes the training process using GPT-4 for knowledge distillation. When both users showed deep interest in this page (when the user pauses scrolling), the system detected {\bf Sustained Attention} and triggered the proactive search process.

As shown in Table~\ref{tab:demo_results},
despite seeing the same text, H-MAPS generated different questions by integrating the current view with the Hierarchical Memory.
For the NLP user, the system interpreted reflection tokens as computational overhead that impacts latency. Capturing terms like ``distilled prompts,'' it generated a query on \textit{inference latency optimization}.
Conversely, for the HCI user, the system interpreted reflection tokens as explainability signals for verification. Inferring that they serve as interface elements, it formulated a query on \textit{trust and transparency}.

\subsection{Retrieval Utility and Verification Support} 
The retrieval results, summarized in Table~\ref{tab:demo_results}, illustrate the system's capacity to steer information seeking based on user profiles.

For the NLP user, the system presented \textit{FedRAG} and \textit{RAG Foundry}. These results are relevant as they provide practical tools to implement and optimize the training processes described on the Self-RAG paper.
For the HCI user, H-MAPS retrieved \textit{Designing Transparency}, which utilizes the 3-gap framework to study information asymmetry. This provides the theoretical foundation needed to evaluate how reflection tokens should be presented to users.

It is important to note that proactive retrieval does not guarantee perfect relevance for every query. However, the key feature of H-MAPS is that it shifts the user's cognitive task from manual search to intent verification. By presenting the explicit question alongside the results, the user can instantly verify whether the assistant’s inferred intent aligns with their own. Even if the retrieved papers are not perfect matches, this explicit feedback loop allows the researcher to maintain their reading flow or quickly dismiss the suggestion, reducing the cognitive cost compared to manual query formulation.

\subsection{System Latency}
We measured the system latency on a desktop computer with a Ryzen9 9950X3D processor and 64GB RAM, without GPU acceleration.
The average end-to-end processing time was approximately 12.2 seconds on average.
The breakdown by component is as follows:
\textbf{(1) Screen Sensing} and \textbf{(2) Hierarchical Memory}: 8.0s;
\textbf{(4) Question Generation}: 2.7s using the cloud-based model; and
\textbf{(5) Local Search}: 1.5s.
Although this latency may be insufficient for interactive ad-hoc retrieval, 
it is suitable for proactive search assistance operating as a background process.

%% file: 5Conclusion.tex
\section{Conclusion}
In this paper, we presented H-MAPS, a proactive search assistant designed to transform the cognitively demanding process of scientific reading into a seamless knowledge-gathering experience. By synthesizing a three-layered hierarchical memory, comprising long-term research interests, session-level context, and immediate on-screen focus, the system provides context-aware search results.
Through a scenario, we showed that H-MAPS effectively disambiguates latent information needs, providing divergent and  relevant literature to researchers with different expertise even when they are engaged with the identical text. 

While our demonstration highlights the potential of H-MAPS, several limitations remain.
First, there is a fundamental throughput asymmetry between the user's reading speed and the local LLM's processing time. When a user scrolls continuously, the rate of newly accumulated text buffer ($B$) outpaces the local LLM's summarization capability (taking approximately 8s per chunk). This bottleneck causes either substantial lags in memory construction or missing reading contexts if unprocessed buffers are discarded, which occasionally destabilizes the \emph{Sustained Attention} trigger.
Second, imperfect relevance and heuristic sanitization require further evaluation. 
In future work, we plan to address this throughput issue by employing smaller, faster models (e.g., distilled models or specialized encoders) for real-time micro-summarization. Furthermore, we will evaluate trigger policies and privacy-utility trade-offs at scale, and explore hybrid retrieval to improve robustness and reduce the heavy local storage footprint.

%% file: 6Acknowledgments.tex
\begin{acks}

This work was supported by JSPS KAKENHI Grant Number \\JP23K28090, and JST PRESTO, Japan, Grant Number JPMJPR25T2.

The authors acknowledge the peoples of the Woi Wurrung and Boon Wurrung language groups of the eastern Kulin Nation on whose unceded lands ACM SIGIR 2026 was hosted. We pay our respects to their Elders past and present, and extend that respect to all Aboriginal and Torres Strait Islander peoples today and their continuing connection to land, sea, sky, and community.
\end{acks}